# Excitonic Photoluminescence properties of nanocrystalline GaSb and $Ga_{0.62}In_{0.38}Sb$ embedded in silica films

Fa-Min Liu [1], Tian-Min Wang [1], Li-De Zhang[2], Guo-Hua Li[3], He-Xiang Han[3]

[1] Center of Material Physics and Chemistry, School of Science, Beijing University of Aeronautics and Astronautics, Beijing 100083, P. R. China
< Tel. +86-10-82317941, Fax. +86-10-82315933, E-mail: fmliu@aphy.iphy.ac.cn   ; OR fmliu@etang.com

[2] Institute of Solid State Physics, Chinese Academy of Sciences, Hefei 230031, P. R. China

[3] National Laboratory for Superlattices and Microstructure, Institute of Semiconductors, Chinese Academy of Sciences, Beijing 100083, P. R. China

**Abstract:** The GaSb and $Ga_{0.62}In_{0.38}Sb$ nanocrystals were embedded in the $SiO_2$ films by radio frequency magnetron co-sputtering and were grown on GaSb and Si substrates at different temperatures. We present results on the 10K excitonic photoluminescence (PL) properties of nanocrystalline GaSb and $Ga_{0.62}In_{0.38}Sb$ as a function of their size. The measurements show that the PL of the GaSb and $Ga_{0.62}In_{0.38}Sb$ nanocrystallites follows the quantum confinement model very closely. By using deconvolution of PL spectra, origins of structures in photoluminescence were identified.

# 1. Introduction

Nanostructures have become a topic of increasing interest because of their interesting optical properties and their potential for applications in optoelectronic devices such as in fast switching and optical signal processing [1-9]. The behavior of semiconductor nanostructures embedded in dielectric host is significantly different from other low-dimensional semiconductor heterostructures [1,2,10-13,17]. It is well known that when the dimensions of crystallites approach the atomic scale, significant changes occur in the electronic, optical, magnetic, or thermodynamic properties compared with those of bulk materials, and also electronic states of these nanostructures have to be treated with an account for interactions with real surfaces or interfaces and the surrounding material. These interactions redistribute the energy relaxation of photoexcited electron-hole pairs and excitons to radiative and nonradiative channels of recombination, which complicates considerably the understanding of the photoluminescence (PL) properties of dielectrically confined quantum well radiations. Among the III-V semiconductor, Gallium antimonide is one of the most promising candidates for applications in electro-optical devices such as detectors, photodiodes, lasers and waveguides over a very broad infrared band. However, so far very little has reported on the PL characterization of GaSb and $Ga_{0.62}In_{0.38}Sb$ nanocrystals containing in the $SiO_2$ matrix. In this paper, we concentrate on the photoluminescence experiment with GaSb and $Ga_{0.62}In_{0.38}Sb$ nanocrystals embedded in silica host. Compared to GaSb bulk, a large blue shift of the maximum of the PL of nanocrystalline GaSb and $Ga_{0.62}In_{0.38}Sb$ has been found at 10 K. The origin of this phenomenon is owed to quantum confinement effect and the strong Coulomb interaction anisotropy in semiconductor insulator quantum well radiations.

## 2. Experimental procedure

The GaSb and $Ga_{0.62}In_{0.38}Sb$ nanocrystals were embedded in the $SiO_2$ films by radio frequency magnetron co-sputtering and were grown on a substrate at different temperatures. The details of the samples prepared and the structures described were published elsewhere [14-17]. The target used was a $SiO_2$ glassy plate 60 mm in diameter, with several chips of polycrystalline GaSb and InSb attached. By varying the numbers of GaSb and InSb chips, the semiconductor concentration in the deposited film can be controlled to a desired value. In this work the effective sputtered area ratio of the semiconductor chips to $SiO_2$ plate was fixed at 20 %. Some glass substrates (10 mm × 10 mm × 1 mm) were ultrasonically washed successively in acetone, alcohol, and deionized water to obtain a clean surface before being placed in a vacuum chamber. The target was separated from the substrate by 5 cm. When the chamber was evacuated to a pressure of $2\times10^{-5}$ Torr high purity (99.9999%) argon gas was introduced. During sputtering, the chamber pressure was maintained at $1.5 \times 10^{-2}$ Torr. The output voltage to the radio frequency sputter gun was 1000 V. All runs started with presputtering the target for 15 minutes to ensure good quality GaSb and $Ga_{1-x}In_xSb$ films.

Photoluminescence (PL) measurements were carried out using a Jobin-Yvon HRD-I double grating monochromator with a microscope and detected by a APD

CSW-202A cooled GaAs photomultiplier operating in photon counting mode. The monochromator step was 0.5 nm and the count time of the detector ranged from 0.5 to 1s. A He-Cd laser operating at 325.0 nm was used as the excitation source with excitation level of 250 W/cm$^2$. A resolution of 0.5 nm was used in our measurements. All PL spectra were recorded at 10 K by attaching the sample in the cold finger of an APD refrigeration system. The measured spectra have been corrected according to the system response.

The mean grain sizes of GaSb and $Ga_{0.62}In_{0.38}Sb$ nanocrystals were evaluated using the X-ray diffraction (XRD) data [15,16] and the Scherrer formula

$$D = 0.9 \lambda / B \cos\theta_B \tag{1}$$

Where $\lambda$, $\theta_B$, and B are the X-ray wavelength (0.154056 nm), Bragg diffraction angle, and full linewidth at half maximum of the diffraction peak, respectively. And, the experimental results [15,16] indicate that the particle size increases with increasing the growing temperature.

**3. Results and discussion**

The spectral position of the 10K excitonic photoluminescence peaks of GaSb nanoparticles depends on the nanocrystal size as shown in Fig. 1. The large blue shift of this band suggests that the band correspond to the electron hole pair recombination where at least one carrier is deeply localized. We can see three blue peaks of PL depending on the size of GaSb nanocrystals. Fig. 1(a) shows the 10K excitonic photoluminescence of nanocrystalline GaSb-SiO$_2$ composite film grown on GaSb substrate at 673 K for 3 hours. The size of GaSb nanocrystals is about 5.3 nm. Three PL peaks in 447.9, 495.5 and 533.8 nm positions come in sight. Fig. 1 (b) and (c) show the 10K excitonic photoluminescence of nanocrystalline GaSb-SiO$_2$ composite films grown on GaSb substrate at 623 K and 573 K for 3 hours, respectively. The sizes of these GaSb nanoparticles are about 4.9 and 4.5 nm, respectively. One sees three PL peaks (438.4 nm, 492.5 nm and 573.3 nm) in Fig. 1 (b) and three PL peaks (432.2nm, 486.0 nm and 535.2 nm) in Fig. 1 (c), respectively. Fig. 2 presents the Gaussian fit and deconvolution of the photoluminescence spectra in Fig. 1 as mentioned above. One can see that the photoluminescence peaks of nanocrystalline GaSb-SiO$_2$ composite films are mainly composed of three PL peaks. The first two peaks are strongly dependent on the size of GaSb nanoparticles. So we confirm that the photoluminescence process take place across the band gap of GaSb nanocrystals. The last PL peak is independent on the size of GaSb nanocrystals. It would be owed to the defect SiO$_2$ film. We further estimate the PL position and discuss the PL structure in the following.

It is well known that the undoped GaSb is usually always p-type semiconductor and has a high concentration of native acceptor [18]. And, GaSb has a band gap of 0.72 eV at room temperature and is a direct band gap material. Under low temperature, GaSb bulk has two main PL peak in the position of 1.5938 μm (0.778 eV) and 1.5576 μm (0.796 eV)[19]. The received PL peaks of GaSb bulk are free excitons from the conduction to valence band transition and, the free excitons from the conduction to neutral acceptor band transition. One can see it in Fig. 3. Apart from these two PL peaks, there are many weak radiative transitions across the band gap.

Obviously, the PL peaks of GaSb nanocrystals have larger blue shifts than that of GaSb bulk from the data mentioned above. These phenomena can be interpreted by the quantum size effect. Firstly, one usually uses Varshni formula [20,21] to describe the variation of semiconductor band gap as temperature under low temperature.

$$E_g(T) = E_g(0) - AT^2/(T+B) \quad (2)$$

Chen et al. [22] studied the PL of GaSb bulk as the function of the temperature varied. They got the following formula for GaSb bulk at low temperature.

$$E_g(T) = 0.812 - 4.2 \times 10^{-4} T^2/(T+140) \quad (3)$$

But, when Wu and Chen [23] fitted their PL data of GaSb bulk at low temperature by using the equation (2), the band gap as a function of temperature they got can be described as follows

$$E_g(T) = 0.813 - [1.08 \times 10^{-4} T^2/(T-10.3)] \quad (4)$$

For GaSb bulk, we consider the band gap as $E_g(10K) = 0.8117$ eV and $E_g(10K) = 0.849$ eV by using the equation (3) and (4), respectively.

Secondly, for GaSb nanocrystals, there are large shifts and broad PL peak according to the quantum confinement theory of Brus and Kayanuma [24-25]. From the predominant point of the quantum confinement theory, the effective Bohr radius $a_B$ of the exciton in the bulk material play an important role in quantum confinement. Usually, $a_B$ can be expressed as

$$a_B = 4\pi\varepsilon_o\varepsilon h^2/4\pi^2\mu e^2 \quad (5)$$

Where $\varepsilon_o$ is vacuum dielectric constant, $\varepsilon$ is relative dielectric constant, h is Planck constant, µ is reduced mass of an electron-hole pair, and e is electronic charge. For GaSb, $\varepsilon$=15.69, µ = 0.041$m_o$, we get its $a_B \approx 20.5$ nm from formula (5). Brus [24] and Kayanuma [25] pointed out: If $a_B$ is much larger than the diameter of nanoparticles (that is $a_B$>>d = 2R, where R is radius of nanoparticles), quantum confinement is obvious, in which the confinement energy dominates over the Coulomb interactions. And also, the confinement enhances the electron-hole-direct and exchange Coulomb interactions because the spatial overlap of the electron and hole wave functions strongly increases with decreasing size. In this work, the particle size of the GaSb semiconductor crystallite ranges from 4 to 6 nm. Therefore the strong quantum confinement leads to a blue-shifted PL peak. According to the effective mass approximation, for a spherical nanoparticle of diameter d, the variation of band gap of semiconductor nanoparticles corresponding to the bulk case is express as

$$\Delta E_g = E - E_g \approx h^2 \pi^2 / 2 d^2 \mu - 1.8 e^2 / 2\pi \varepsilon \varepsilon_0 d + \text{smaller terms} \qquad (6)$$

Where µ is a reduced mass of an electron-hole pair, $\varepsilon_0$ and $\varepsilon$ are the dielectric constants of vacuum and the semiconductor, h is Planck's constant, and e is electronic charge. From the equation (3), (4) and (6), the band gap energy shifts of PL peaks as the sizes of nanocrystals are shown in Fig. 4. Obviously, according to equation (3) of Chen, et al. [22], and equation (4) of Wu, et al. [23], we have got the approximately linear curve for PL peak α and PL peak β of nanocrystalline GaSb as a function of $d^{-2}$. These show that the experimental data is basically agreement with the quantum confinement theory. So, one can confirmed that the optical transitions in GaSb nanocrystals is in the Brillouin Zone (BZ) and is of the direct-band gap character. The prominent structures of PL peak α and PL peak β are the free excitons from conduction to valence and from conduction to neutral acceptor level transition, respectively. The other PL peak (about 535 nm) is originated from the $SiO_2$ film and other factor. The broad photoluminescence peaks observed in Fig. 1 are attributed to synthesis of the three PL peaks and the size fluctuations of GaSb nanoparticles. First, there is a distribution of GaSb nanoparticles in the $SiO_2$ films. One can get a broad PL peaks for a GaSb nanoparticles according to the quantum size effect. The very broad PL is the statistic iterative results of PL of the size fluctuations of GaSb nanoparticles. Second, other factors such as defect, impurity and lattice mismatch, lead to widen the band tail. So, these properties of GaSb nanocrystals embedded in $SiO_2$ films is different from the luminescence in epitaxially growing GaSb quantum dots, self-organized in a GaAs matrix using molecular beam epitaxy (MBE) [26], which originates from radiative recombination of 0D holes and spatially separated electrons attracted to them via Coulomb interaction. The GaSb/GaAs system presents an excellent model to study optical properties of type-II quantum wells.

Fig. 5 shows the 10K excitonic photoluminescence of nanocrystalline GaSb-$SiO_2$ composite films grown on the GaSb and Si substrates. Fig. 4 (a) shows the 10K excitonic photoluminescence of nanocrystalline GaSb-$SiO_2$ composite films grown on the GaSb substrate. There are three PL peaks in the position of 444.7, 495.5, and 533.6 nm. Fig. 4 (b) shows the 10K excitonic photoluminescence of nanocrystalline

GaSb-SiO$_2$ composite films grown on the Si substrate. We can also see three PL peaks in the position of 432.1, 493.8, and 533.6 nm. The two films were grown at the same condition (673 K, 3h) and the sizes of GaSb nanocrystals in the films are about 5.3 nm. Compared to Fig. 5 (a) and Fig. 5 (b), we have found that there are more little blue shifts of the films grown on Si substrate than that of the films grown on GaSb. This is because there is a inner stress or strain in GaSb-SiO$_2$/Si films. The lattice parameter of GaSb is 0.6094 nm, and that of Si is 0.5431 nm. The lattice mismatch between GaSb and Si is about 10-12 %. In addition, since the thermal expansion coefficient of GaSb ($6.8 \times 10^{-6}$ K$^{-1}$) is much larger than that of Si ($2.6 \times 10^{-6}$ K$^{-1}$ and silica glass ($0.55 \times 10^{-6}$ K$^{-1}$), GaSb nanoparticles contract more than the silica glass upon cooling. The mismatch of the thermal expansion coefficients and the lattices causes a tensile stress to be exerted upon the GaSb nanoparticles, which causes the lattice constant to distort, leading to a small blue shift of the optical absorption edge. Therefore, the PL peaks of the GaSb-SiO$_2$ composite film grown on Si substrate have tiny larger blue shifts than those of the film grown on GaSb.

The comparison to 10K excitonic photoluminescence of nanocrystalline GaSb-SiO$_2$ and Ga$_{0.62}$In$_{0.38}$Sb-SiO$_2$ composite films are shown in Fig. 6. One can see that the band gap of nanocrystalline Ga$_{0.62}$In$_{0.38}$Sb has some different from that of nanocrystalline GaSb. From Fig. 6, one sees that there is only one PL peak in the position of 479.6 nm for the film of nanocrystalline Ga$_{0.62}$In$_{0.38}$Sb-SiO$_2$ composite film, and the full width at half maximum (FWHM) is widened. This phenomenon is agreement with that of Basu's [27]. It is well known that the In element doped in GaSb, it will be decreased the concentration of native acceptor. Thus, the probability of free excitons recombination with native acceptor excitons is very little.

The 10K excitonic photoluminescence of nanocrystalline $Ga_{0.62}In_{0.38}Sb$-$SiO_2$ composite films grown on the GaSb substrate are shown in Fig. 7. We can see only one blue peak of PL, which depends on the size of $Ga_{0.62}In_{0.38}Sb$ nanocrystals. Fig. 7 (a) shows the 10K excitonic photoluminescence of nanocrystalline $Ga_{0.62}In_{0.38}Sb$-$SiO_2$ composite film grown on GaSb substrate at 673 K for 3 hours. The size of $Ga_{0.62}In_{0.38}Sb$ nanocrystals is about 5.6 nm. Only one PL peaks in 479.6 nm position come in sight. Fig. 7(b) and 7(c) show the 10K excitonic photoluminescence of nanocrystalline $Ga_{0.62}In_{0.38}Sb$-$SiO_2$ composite films grown on GaSb substrate at 623 K and 573 K for 3 hours, respectively. The sizes of these $Ga_{0.62}In_{0.38}Sb$ nanoparticles are about 5.2 and 4.8 nm, respectively. One can also see only one PL peak (460.9 nm) in Fig. 7(b) and one PL peak (444.9 nm) in Fig. 7(c), respectively. Owing to In element doping the GaSb, the concentration of native acceptor is decreased. So many PL peaks were not observed because the probability of free excitons transfer from conductivity to neutral acceptor excitons is sharply decreased. However, deep states are only slightly affected by the confinement process, due to the localization of their electronic wave function. This means that this PL is connected with the recombination of the electron hole pair in which one of the carriers is deeply trapped and the other one is in the nanocrystals interior state or in a shallow trap.

On the other hand, according to the quantum confinement theory of Brus and Kayanuma [24-25] mentioned above, one can well comprehend a large blue PL peak of nanocrystalline $Ga_{0.62}In_{0.38}Sb$ corresponding to its bulk as a function of the size of $Ga_{0.62}In_{0.38}Sb$ nanocrystals.

## 4. Conclusions

In conclusion, we have investigated the excitonic photoluminescence of nanocrystalline GaSb and $Ga_{0.62}In_{0.38}Sb$ embedded in the $SiO_2$ films by radio frequency magnetron co-sputtering. The excitonic photoluminescence peaks of the nanocrystalline GaSb and $Ga_{0.62}In_{0.38}Sb$ have very larger blue shifts than that of bulk GaSb and $Ga_{0.62}In_{0.38}Sb$. These phenomena have been well explained by quantum confinement theory. And the PL mechanism of nanocrystalline GaSb and $Ga_{0.62}In_{0.38}Sb$ has been discussed.

**Acknowledgments**

We would like to thank Prof. Y. L. Liu (Institute of Physics, Chinese Academy of Sciences) for helping to do the Gaussian fit and deconvolution of our PL spectra. This work was supported by The National Climbing Program of China and in part by the NSF of China under Contract No. 59982002 and Post-doctoral Science Foundation of China.

**Captions for Figures:**

**Fig.1** 10K excitonic photoluminescence of nanocrystalline GaSb-SiO$_2$ composite films grown on the GaSb substrate   (a) 673 K-3h, 5.3 nm   (b) 623 K-3h，4.9 nm   (c) 573 K-3h, 4.5 nm

**Fig. 2** Gaussian fit and deconvolution process of 10 K 10K excitonic photoluminescence of nanocrystalline GaSb-SiO$_2$ composite films. Fig. 2 (a), (b) and (c) correspond to Fig. 1 (a), (b) and (c) respectively.

**Fig.3**   The principle figure of electron-hole pair recombination across the band gap. Ec → Ev is the band to band transition, and Ec→A is the free excitons-neutral acceptor excitons transition.

**Fig.4**   The band gap energy shifts of PL peaks Vs the sizes of nanocrystals. (a) according to equation (2) of Chen, et al. [22], (b) according to equation (3) of Wu, et al. [23] .

**Fig.5** 2 10K excitonic photoluminescence of nanocrystalline GaSb-SiO$_2$ composite films grown on the different substrate

**Fig.6** 10K excitonic photoluminescence of nanocrystalline GaSb-SiO$_2$ and Ga$_{0.62}$In$_{0.38}$Sb-SiO$_2$ composite films

**Fig.7** 10K excitonic photoluminescence of nanocrystalline Ga$_{0.62}$In$_{0.38}$Sb−SiO$_2$ composite films grown on the GaSb substrate   (a) 673 K-3h, 5.6 nm   (b) 623 K-3h，5.3 nm   (c) 573 K-3h, 4.9 nm

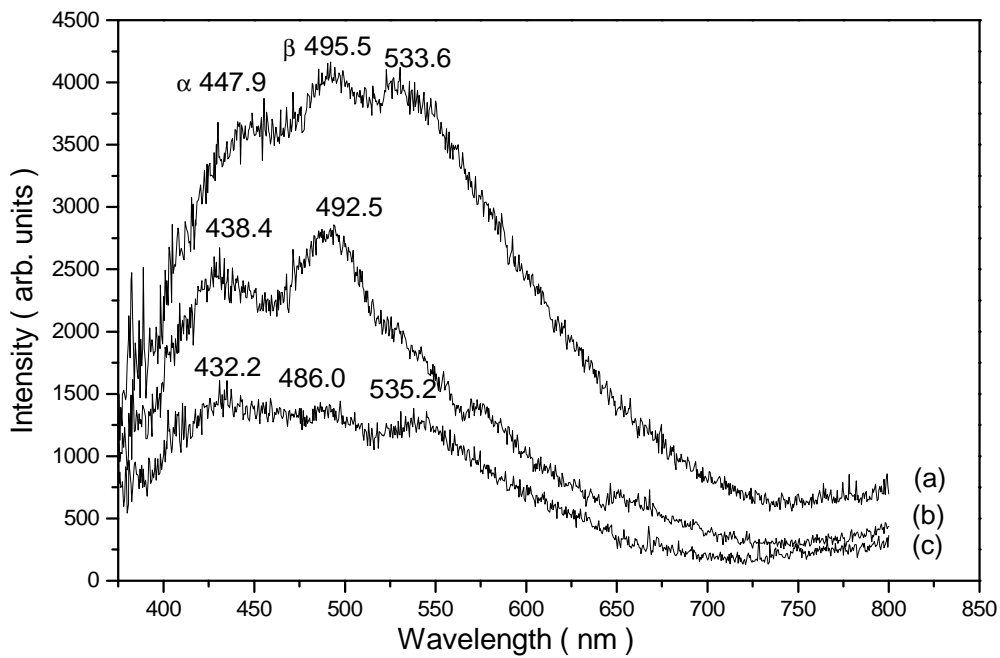

Fig. 1 10 K excitonic photoluminescence of nanocrystalline GaSb-SiO$_2$ composite films grown on the GaSb substrate (a) 673 K-3h, 5.3 nm (b) 623 K-3h, 4.9 nm (c) 573 K -3h, 4.5 nm

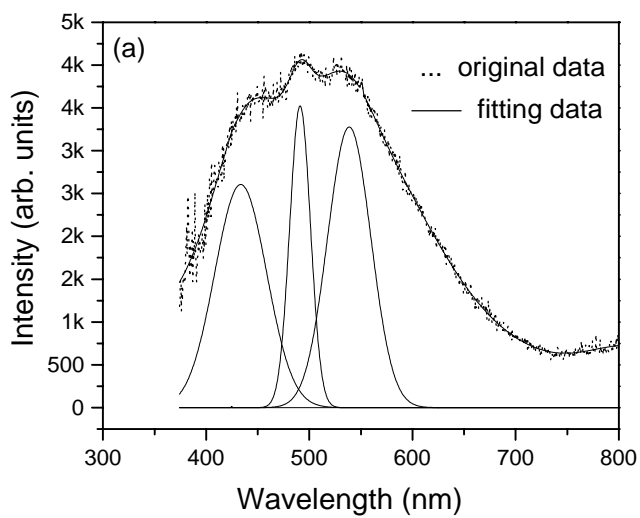

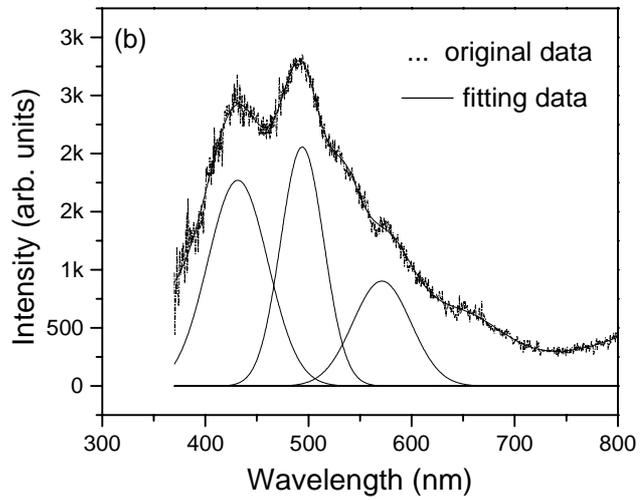

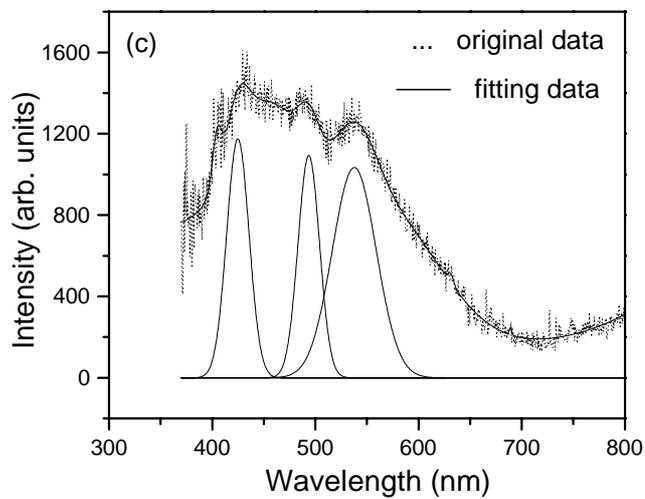

Fig. 2 Gaussian fit and deconvolution process of 10 K excitonic photoluminescence of nanocrystalline GaSb-SiO$_2$ composite films. Fig. 2 (a), (b) and (c) correspond to Fig. 1 (a), (b) and (c) respectively.

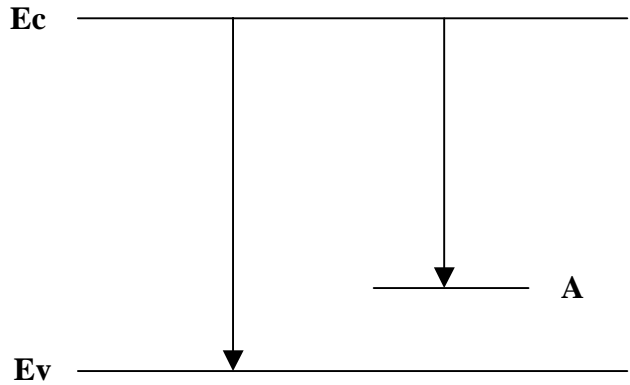

Fig.3  The principle figure of electron-hole pair recombination across the band gap. Ec → Ev is the band to band transition, and Ec →A is the free excitons-neutral acceptor excitons transition.

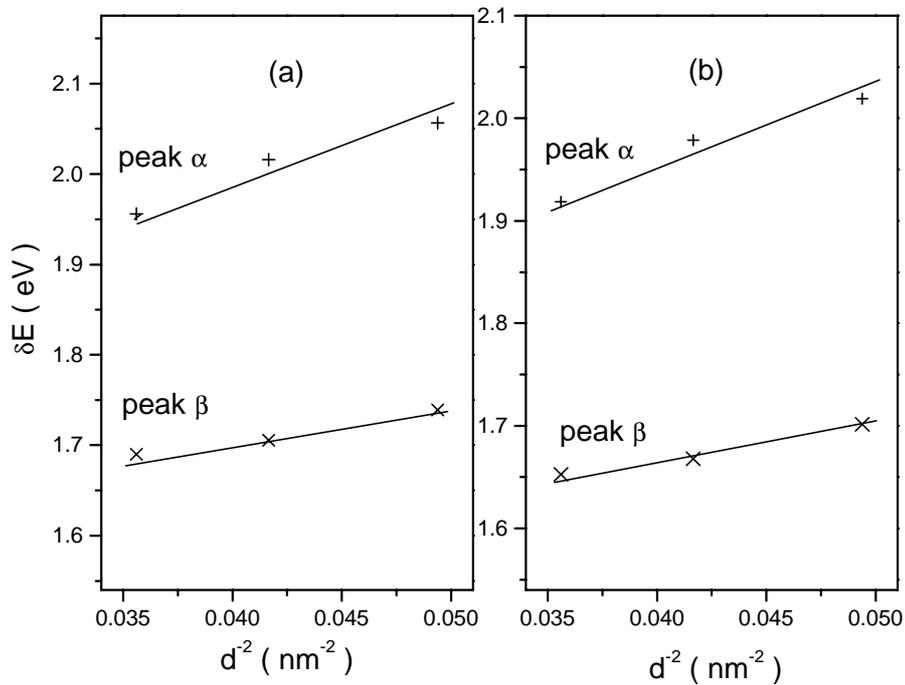

Fig.4  The band gap energy shifts of PL peaks Vs the sizes of nanocrystals. (a) according to equation (2) of Chen, et al. [22], (b) according to equation (3) of Wu, et al.

[23] .

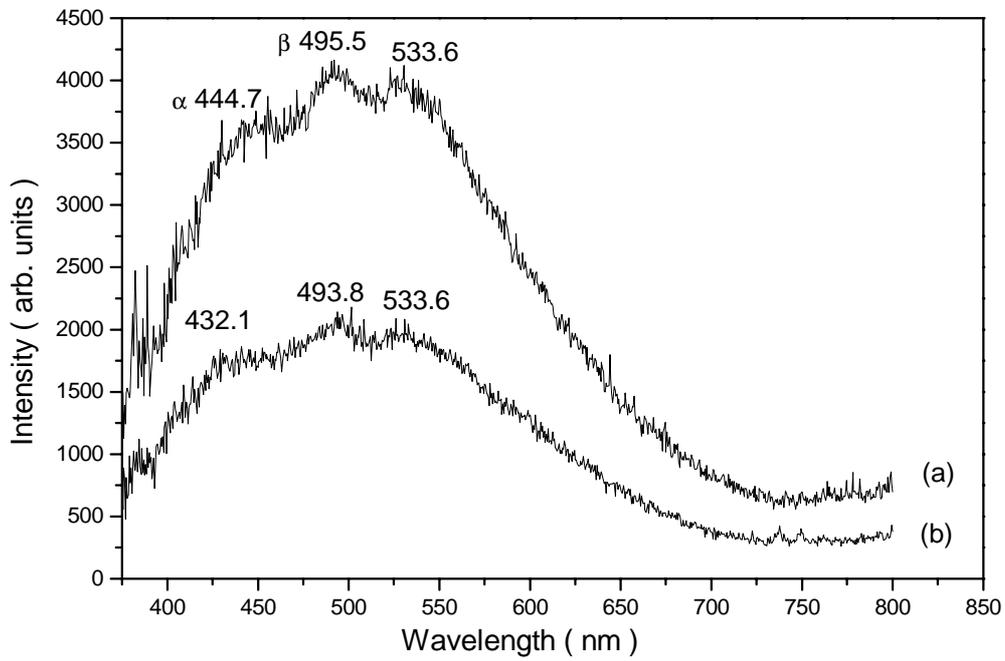

Fig. 5 10 K excitonic photoluminescence of nanocrystalline GaSb-SiO$_2$ composite films grown on the different substrate

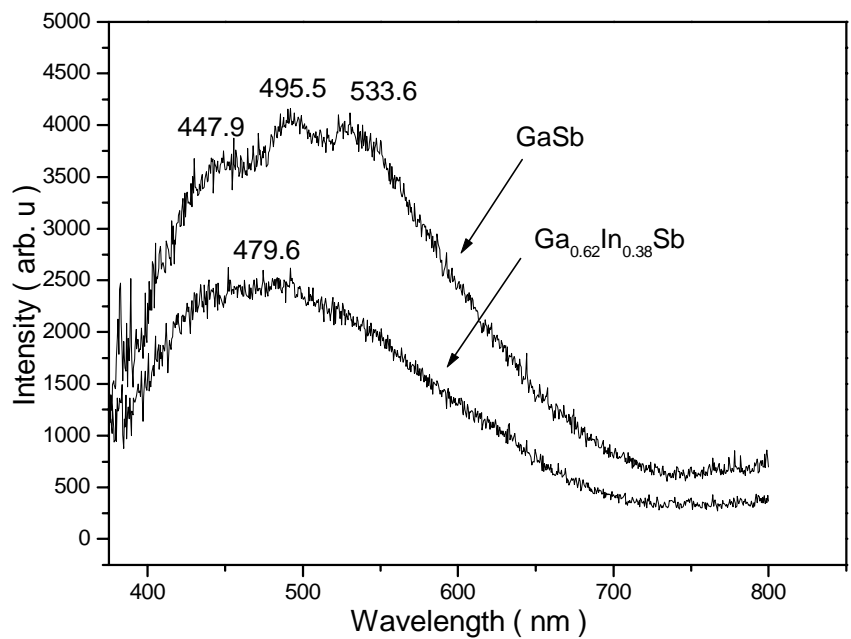

Fig. 6 10 K excitonic photoluminescence of nanocrystalline GaSb-SiO$_2$ and Ga$_{0.62}$In$_{0.38}$Sb-SiO$_2$ composite films

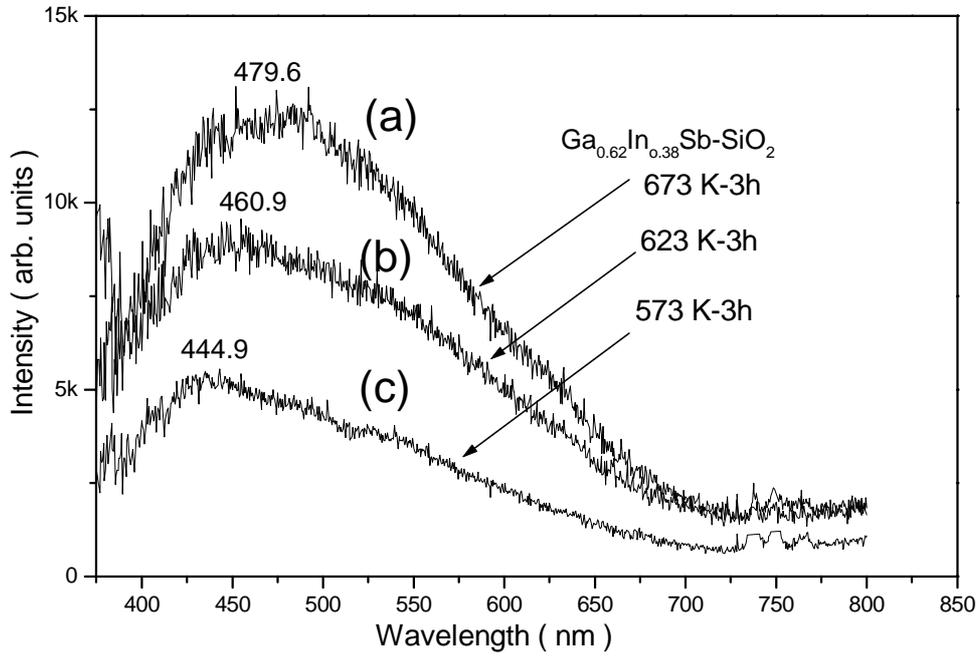

Fig. 7  10 K excitonic photoluminescence of nanocrystalline $Ga_{0.62}In_{0.38}Sb-SiO_2$ composite films grown on the GaSb substrate    (a) 673K-3h, 5.6 nm    (b) 623 K-3h, 5.2 nm    (c) 573 K-3h, 4.8 nm